\documentclass[%
 reprint,
superscriptaddress,
nofootinbib,
 amsmath,amssymb,
 aps,
 prl,
floatfix,
]{revtex4-2}

\usepackage{enumitem}

\usepackage{graphicx}
\usepackage{dcolumn}
\usepackage{bm}

\usepackage{epsfig}
\usepackage{amsmath}
\input{epsf}
\usepackage{psfrag}
\usepackage{xcolor}
\usepackage{pdfpages}
\usepackage[normalem]{ulem}
\makeatletter
\AtBeginDocument{\let\LS@rot\@undefined}
\makeatother
\newcommand{\bea}{\begin{eqnarray}}
\newcommand{\eea}{\end{eqnarray}}

\newcommand{\nn}{\nonumber}

\begin{document}

\title{Illuminating Nucleon Gluon Interference via Calorimetric Asymmetry}

\author{Xiao Lin Li}%
\affiliation{Center of Advanced Quantum Studies, Department of Physics, Beijing Normal University, Beijing, 100875, China}

\author{Xiaohui Liu}
 \email{xiliu@bnu.edu.cn}
 \affiliation{Center of Advanced Quantum Studies, Department of Physics, Beijing Normal University, Beijing, 100875, China}
 \affiliation{Key Laboratory of Multi-scale Spin Physics, Ministry of Education, Beijing Normal University, Beijing 100875, China}
 \affiliation{Center for High Energy Physics, Peking University, Beijing 100871, China}

\author{Feng Yuan}%
 \email{fyuan@lbl.gov}
\affiliation{Nuclear Science Division, Lawrence Berkeley National Laboratory, Berkeley, CA 94720, USA}%

\author{Hua Xing Zhu}%
 \email{zhuhx@pku.edu.cn}
\affiliation{School of Physics, Peking University, Beijing 100871, China}%
\affiliation{Center for High Energy Physics, Peking University, Beijing 100871, China}

\begin{abstract}
We present an innovative approach to the linearly polarized gluons confined inside the unpolarized nucleon in lepton-nucleon scattering. Our method analyzes the correlation of energy flows at azimuthal separations $\phi$.
The interference of the spinning gluon with both positive and negative helicities translates into a $\cos(2\phi)$ asymmetry imprinted on the detector. Unlike the conventional transverse momentum dependent (TMD) probes, the $\cos(2\phi)$ asymmetry in this approach is preserved by rotational symmetry, holds to all orders, and is free of radiation contamination, thus expected to provide the exquisite signature of the nucleon linearly polarized gluons. 

\end{abstract}

\maketitle

\begin{figure}[htbp]
  \begin{center}
   \includegraphics[scale=0.46]{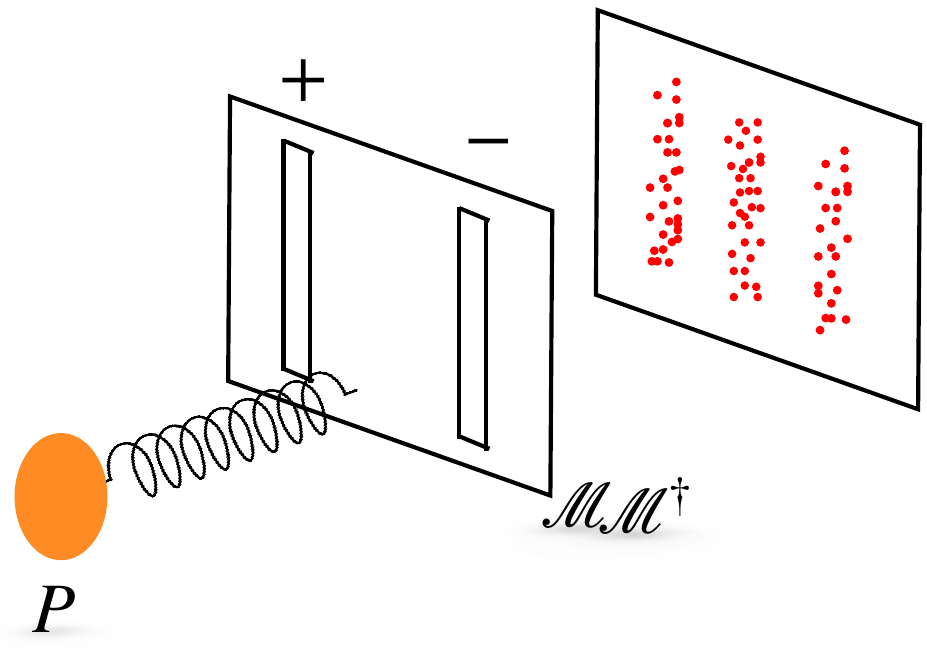} 
\caption{Nucleon structure as a double split experiment in the helicity space.}
  \label{fg:helicity}
 \end{center}
\end{figure}

\textbf{\textit{  Introduction.}} Quarks and gluons that are confined within
nucleons will be examined in unprecedented detail 
at the next generation QCD facilities~\cite{Accardi:2012qut, AbdulKhalek:2021gbh, Proceedings:2020eah}. 
Extracting their fundamental properties requires the analysis of scattering data to reveal their distributions inside nucleons. It is now widely recognized that even within an unpolarized nucleon, partons can exhibit polarization, leading to a scattering cross-section of the schematic form
\bea 
\sigma  &\propto & 
|\hat{{\cal M}}|+\rangle  + \hat{{\cal M}} |-\rangle |^2  \nn \\ 
&=&{\sum_{i=+,-} }\langle i|
\hat{{\cal M}}^\dagger \hat{{\cal M}}  |i\rangle  
+ \big(\langle + |\hat{{\cal M}}^\dagger \hat{{\cal M}}  |-\rangle 
+ c.c. \big)\,,
\eea 
in which $|i\rangle$ denotes the helicity state of the parton out of the hadron, and $\hat{{\cal M}}$ is the transition operator. 
Thus far experimental and theoretical studies have placed extensive focus on the first trace term, which brings about the most familiar unpolarized parton distributions, such as the collinear parton distribution functions (PDFs) and the unpolarized transverse momentum-dependent PDFs (TMDs). While these unpolarized distributions have provided us with valuable insights into the dynamics of strong interactions, the off-diagonal terms contain the intrinsic quantum effects. The operator $\hat{{\cal M}}$ acts as a screen in helicity space, leading to a double-slit interference phenomenon when both $i = +$ and $i = -$ are allowed, as illustrated in Fig.~\ref{fg:helicity}. A similar effect in the context of final states has drawn recent discussions in jet physics~\cite{Chen:2020adz, Chen:2021gdk, Karlberg:2021kwr} and top physics~\cite{Yu:2021zpe}, leading to an interesting application of 2D conformal symmetry in 4D collider physics~\cite{Chen:2022jhb,Chang:2022ryc}. 

The knowledge of the off-diagonal 
contribution requires the quantum description of a nucleon by the density matrix $\rho_{ij} = |i\rangle \langle j|$, with $i,j=+/-$, where $
\frac{1}{2}
{\rm Tr}\rho$ gives the unpolarized distributions. 
Out of that, the entropy of a nucleon in the helicity space can also be defined, $S = -\rho \ln \rho$. Furthermore, the concepts, such as the positivity of $\rho$ and the maximum entropy principle, may be introduced and tested in the hadron structure studies. 
In the following, we focus on the hadron structure associated with the gluon distribution, and we refer to the off-diagonal contribution as the linearly polarized gluon distribution. 

In order to observe the effect, one has to introduce a transverse reference direction that goes beyond the conventional collinear PDFs. In the literature, this has been demonstrated in the generalized parton distribution (GPD) framework~\cite{Ji:1996ek, Muller:1994ses, Ji:1996nm, Radyushkin:1997ki} and the TMD framework. In the GPD formalism, the associated distribution is also called the transversity or helicity-flip gluon GPD~\cite{Diehl:1997bu,Hoodbhoy:1998vm,Belitsky:2000jk}, where the momentum transfer between the initial and final state hadrons plays the role as a reference direction. Similarly, in the TMD formalism, the transverse momentum of the gluon helps to define the linearly polarized gluon distribution~\cite{Boer:1997nt,Mulders:2000sh}. An anticipated outcome of this novel gluon distribution is a $\cos(2\phi)$ azimuthal angular asymmetry in the associated hard scattering processes~\cite{Diehl:1997bu,Hoodbhoy:1998vm,Belitsky:2000jk,Boer:2010zf, Metz:2011wb, Pisano:2013cya, Hatta:2020bgy,Hatta:2021jcd,Esha:2022ovp,Caucal:2023fsf}. While, in the 
TMD framework, the asymmetry could receive additional $\cos(n\phi)$  corrections not associated with the nucleon target's parton polarization~\cite{Hatta:2020bgy,Hatta:2021jcd,Caucal:2023fsf}, which overshadows the naive $\cos(2\phi)$ expectation.

In this paper, we apply recently proposed nucleon energy-energy correlator (NEEC)~\cite{Liu:2022wop}, which is a novel extrapolation of the energy-energy correlator~\cite{Basham:1978bw} from final-state jet substructure~\cite{Chen:2020vvp,Komiske:2022enw} to the nucleon structure, to introduce a fresh strategy to identify the linearly polarized gluon distribution in an unpolarized hadron. One standout feature 
of our proposal is that the gluon operator in the NEEC follows the (helicity-dependent) Dokshitzer-Gribov-Lipatov-Altarelli-Parisi (DGLAP) evolution~\cite{Cao:2023oef} and a collinear factorization is applied to compute the associated differential cross sections and the relevant azimuthal angular asymmetries. This contrasts heavily with the linearly polarized gluon distribution in the TMD formalism, where soft gluon radiation plays an important role. The customary approach to the linearly polarized gluon involves the gluonic TMD distributions by observing the $\cos(2\phi)$ asymmetry between the slight imbalance ${\vec k}_T$ and the leading jet momentum ${\vec P}_{J,T}$ in di-jet/di-hadron production in the DIS process~\cite{Boer:2010zf, Metz:2011wb, Pisano:2013cya, Hatta:2020bgy,Hatta:2021jcd}. However, soft radiations that are unrelated to the nucleon target's partons can also generate significant anisotropy in the form of $\cos(n\phi)$ (where $n=1,2, \dots$)~\cite{Hatta:2020bgy,Hatta:2021jcd,Caucal:2023fsf} and thus obscure the physics. Additionally, Sudakov logarithms that lead to a significant suppression in the non-perturbative region~\cite{Boer:2017xpy} complicates the analysis of other TMD probes such as Higgs production~\cite{Boer:2011kf,Gutierrez-Reyes:2019rug} in hadronic collisions.

Unlike its predecessors, the linearly polarized gluon contribution in the NEEC is formulated in the collinear factorization for the inclusive process, without the contamination from soft radiation, eliminating the need for the Sudakov resummation. The $\cos(2\phi)$ signature is preserved by rotational symmetry and persists to all orders. Therefore this methodology provides a unique opportunity to observe the interference effects from the spinning gluon inside the hadrons. 



\textbf{\textit{Helicity dependent NEEC.}}
The NEEC was introduced in~\cite{Liu:2022wop} as a new quantity for the nucleon structures, which complements the TMDs and has been demonstrated as an efficient portal to the onset of gluon saturation~\cite{Liu:2023aqb}. The operator definition of the unpolarized NEEC can be found in~\cite{Liu:2022wop,Cao:2023oef}, which involves the asymptotic energy flow operator $\hat{{\cal E}}(\theta_a)$ that records the energy deposition in the calorimeter at a fixed angle $\theta_a$, normalized to the proton energy, but with the azimuthal position $\phi_a$ integrated over. 

If we keep the azimuthal dependence and measure the energy flow into the solid angle $(\theta_a,\phi_a)$, the related flow direction $n_a^\alpha = (1,\sin\theta_a \cos\phi_a, \sin\theta_a \sin \phi_a, \cos\theta_a)$ supplies a chance to map out the intrinsic Lorentz structure of the gluon field in terms of the helicity dependent NEEC,  

\bea\label{eq:fgmunu}
&& f^{\alpha\beta}_{g,{\rm EEC}} (x,\vec{n}_a)
= 
\int \frac{dy^-}{4\pi xP^+ } e^{- i x P^+ \frac{y^-}{2} }   \nn \\ 
&&
\hspace{4.ex} 
\times  
\langle P| 
{\cal F}^{+\alpha}
\left(y^- \right) 
{\cal L}^\dagger[\bm{\infty},y^-]
\hat{{\cal E}}({\vec n}_a)   
{\cal L} [\bm{\infty},0]
{\cal F}^{+\beta}
(0)  |P \rangle  \,
\nn \\ 
&&
= -g_T^{\alpha\beta} f_{g,{\rm EEC}}
+ \left(\frac{n_{a,T}^\alpha n_{a,T}^\beta}{n_{a,T}^2} - \frac{g_{T}^{\alpha\beta}}{2} 
\right) 
d_{g,{\rm EEC}}\,, 
\eea 
where the first equation furnishes the operator definition of the helicity-dependent gluon NEEC in which ${\cal F}$ is the gauge field strength tensor, and ${\cal L}$ is the gauge link. If we average the gluon helicity, we recover the unpolarized gluon NEEC~\cite{Cao:2023oef}. 
In the second equation, $g_T^{\alpha\beta} = g^{\alpha\beta} - \frac{P^\alpha{\bar n}^\beta +  {\bar n}^\alpha P^\beta}{{\bar n}\cdot P}$, with ${\bar n}\cdot P = P^0+P^z \equiv P^+$, $\vec{n}_a = \sin\theta_a (\cos\phi_a,\sin\phi_a)$ and $n_{a,T}^\alpha = (0,\vec{n}_a,0)$ is the transverse component of the light ray vector $n_{a}^\alpha$. The second equation is the most general parameterization of $f_{g,{\rm EEC}}^{\alpha\beta}$ to satisfy rotational covariance around the $z$-axis. 

The coefficient $f_{g,{\rm EEC}}(\theta_a)$ is the unpolarized NEEC~\cite{Cao:2023oef}, while $d_{g,{\rm EEC}}(\theta_a)$ is the {\it linearly polarized gluon NEEC} originated from the interference between different helicity states. To see this, we parameterize the gluon polarization vectors as 
$ 
\epsilon_{\pm}^{\ast\alpha}  
=
\frac{1}{\sqrt{2}   }
\left( 0 \,,  
  1\,, \mp i     \,, 0\right )    
$, 
it is then straightforward to check that 
$ 
 \epsilon_{\pm, \alpha} \epsilon^\ast_{\pm,\beta} f_{g,{\rm EEC}}^{\alpha\beta} = f_{g,{\rm EEC}}
 $, and 
$ \epsilon_{\mp, \alpha} \epsilon^\ast_{\pm,\beta} f_{g,{\rm EEC}}^{\alpha\beta}  = \frac{1}{2}e^{\mp 2i\phi_a} d_{g,{\rm EEC}} 
$, 
which manifests that the linearly polarized gluon NEEC is a consequence of helicity interference. Since the energy flow measurement $\hat{E}({\vec n}_a)$ is isotropic in the azimuthal plane, the nontrivial $\phi_a$ dependence of the $f_{g,{\rm EEC}}^{\alpha\beta}$ probes directly the polarization of the gluon field inside the nucleon.

When $P^+ \theta_a \gg \Lambda_{\rm QCD}$, the NEEC can be further matched onto the collinear PDFs. At ${\cal O}(\alpha_s)$, the matching of the unpolarized NEEC can be found in ~\cite{Cao:2023oef}. The linearly polarized gluon NEEC is calculated from the polarized splitting function and found to be 
\bea 
&&
d_{g,{\rm EEC}}(x,\theta_a^2)  
= 
\frac{\alpha_s }{4\pi^2}
\frac{2}{\theta_a^2} 
\int \frac{dz}{z}  
(1-z) \frac{1-z}{z}   \nn \\ 
&&  \hspace{10.ex}
\times
\frac{x}{z} 
 \left[ C_F
f_q\left(\frac{x}{z} \right) 
+  C_A f_g\left(\frac{x}{z} \right)
\right]  \,, 
\eea 
Here we have averaged over the initial parton color and spin. The evolution of $f_{g,{\rm EEC}}$ follows the DGLAP evolution~\cite{Cao:2023oef}. The $d_{g,{\rm EEC}}$ obeys the helicity-dependent DGLAP equation that resums logarithms $\alpha_s^n \ln^{n-1} \theta_a/\theta_a$ and will be carried out in future work.



\begin{figure}[htbp]
  \begin{center}
   \includegraphics[scale=0.42]{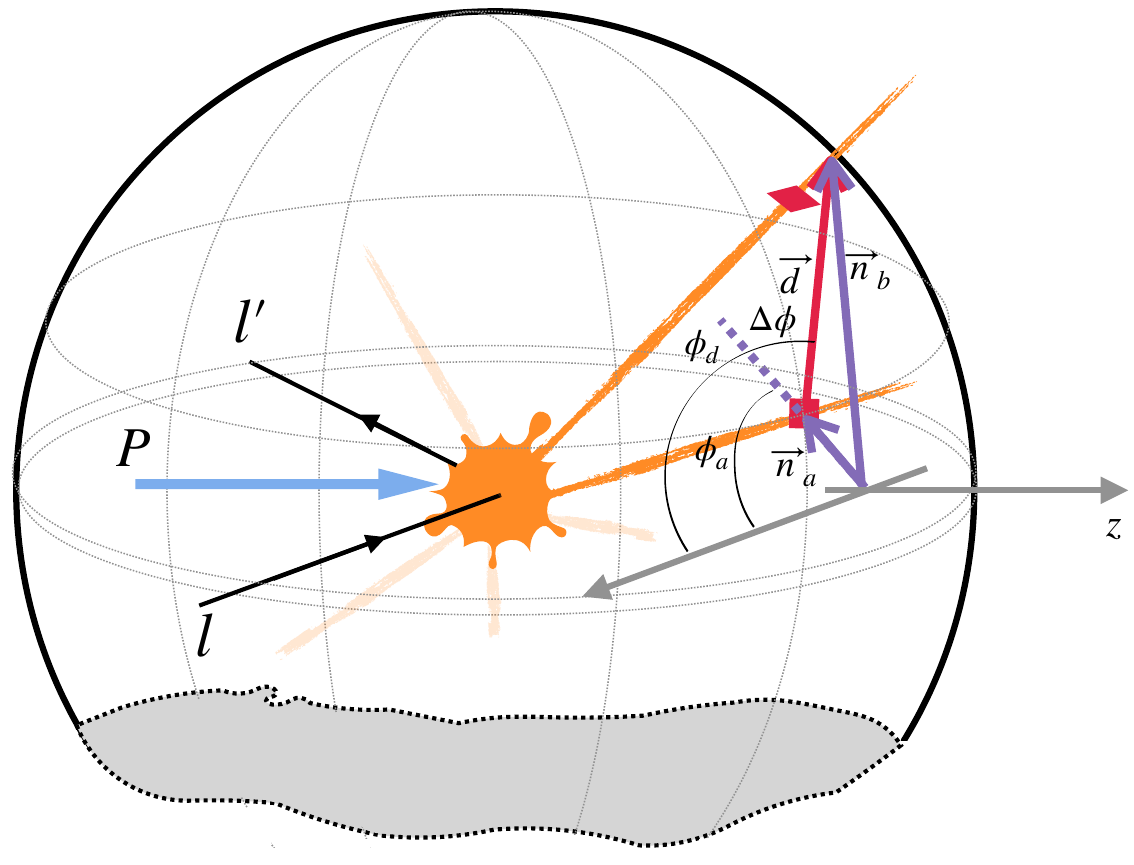} 
\caption{The measurement proposed as a probe of the gluon polarization in the DIS process. The energy flow into different pixels (in red blocks) at ${\vec n}_a$ and ${\vec n}_b$ are recorded, for $\theta_a \ll \theta_b$. $\phi$ angles are measured from the plane where lies the leptons. 
The measurement of $E_i({\vec n}_a)$ induces the NEEC.  }
  \label{fg:measure}
 \end{center}
\end{figure}

\textbf{\textit{Measurement of the Energy Correlator.}}
We consider the unpolarized DIS process in the Breit Frame, in which the incoming proton is along the $z$-axis and the virtual photon generates no transverse momentum with its momentum $q^\mu = (0,0,0,-Q)$. We measure the energy flows that deposit in $2$ arbitrary pixels on the calorimeter located at ${\vec n}_a =\sin\theta_a(\cos\phi_a,\sin \phi_a)$ and ${\vec n}_b = \sin\theta_b(\cos\phi_b,\sin\phi_b)$. Here, $\theta$'s and $\phi$'s are polar and azimuthal angles, respectively. The polar angles are measured with respect to the $z$-axis and the azimuthal angles are measured from the plane spanned by the proton and the leptons, as shown in Fig.~{\ref{fg:measure}}. We then construct $\vec{d} = \vec{n}_b - \vec{n}_a$. 

We require one of the pixels much closer to the proton beam axis, i.e., suppose it is $a$,  then $\theta_a \ll \theta_b$ and when $Q\theta_a \sim {\cal O}(\Lambda_{\rm QCD})$ we probe the NEEC of the proton~\cite{Liu:2022wop}. The other pixel, suppose it is $b$,  is in the central region. Since $\theta_a \ll \theta_b$, the measurement of the energy flow along ${\vec n}_b$ guarantees the inclusive di-jet configuration in the central region to balance the transverse momentum.  

To probe the interference of the gluon helicities, we look at the azimuthal angle difference $ \phi = \phi_d - \phi_a$  between ${\vec n}_a$ and ${\vec d}$~\footnote{One can also measure the azimuthal difference $ \phi'$ between ${\vec n}_b$ and ${\vec n}_a$. The difference between $\phi$ and $\phi'$ vanishes as $\theta_a /\theta_b \to 0$. However, the power correction to the factorization in Eq.~(\ref{eq:sig-z}) could be significant if we use $\phi_b-\phi_a$ directly. A similar strategy is used to suppress the power correction in~\cite{Chen:2021gdk}.}, see Fig.~\ref{fg:measure}. We note that when $\theta_a \to 0$, ${\vec d} \to \vec{n}_b$ and $\phi \to \phi_b-\phi_a$.  
More specifically, we measure the energy-weighted cross-section
\bea\label{eq:e3c} 
&& \Sigma(x_B, Q^2,\cos\theta_{a,b},\phi) \nn \\ 
&=& \sum_{ij}\int  d\sigma(x_B,Q^2) \frac{E_i}{E_P} \frac{E_j}{E_P}
\delta(\vec{n}_a-\vec{n}_i)\delta(\vec{n}_b-\vec{n}_j) \nn \\ 
&& 
\hspace{7.ex}
\times {\cal F}(\phi; \vec{n}_{a,b})\,,
\eea 
where ${\cal F}(\phi; \vec{n}_{a,b})$ imposes the phase space measurement to construct $\phi$.
Here we note that we integrated over the azimuthal angles of the lepton, and the $\phi_{a,b}$. The only azimuthal angle we observe in this measurement is $\phi$.

\begin{figure}[htbp]
  \begin{center}
   \includegraphics[scale=0.29]{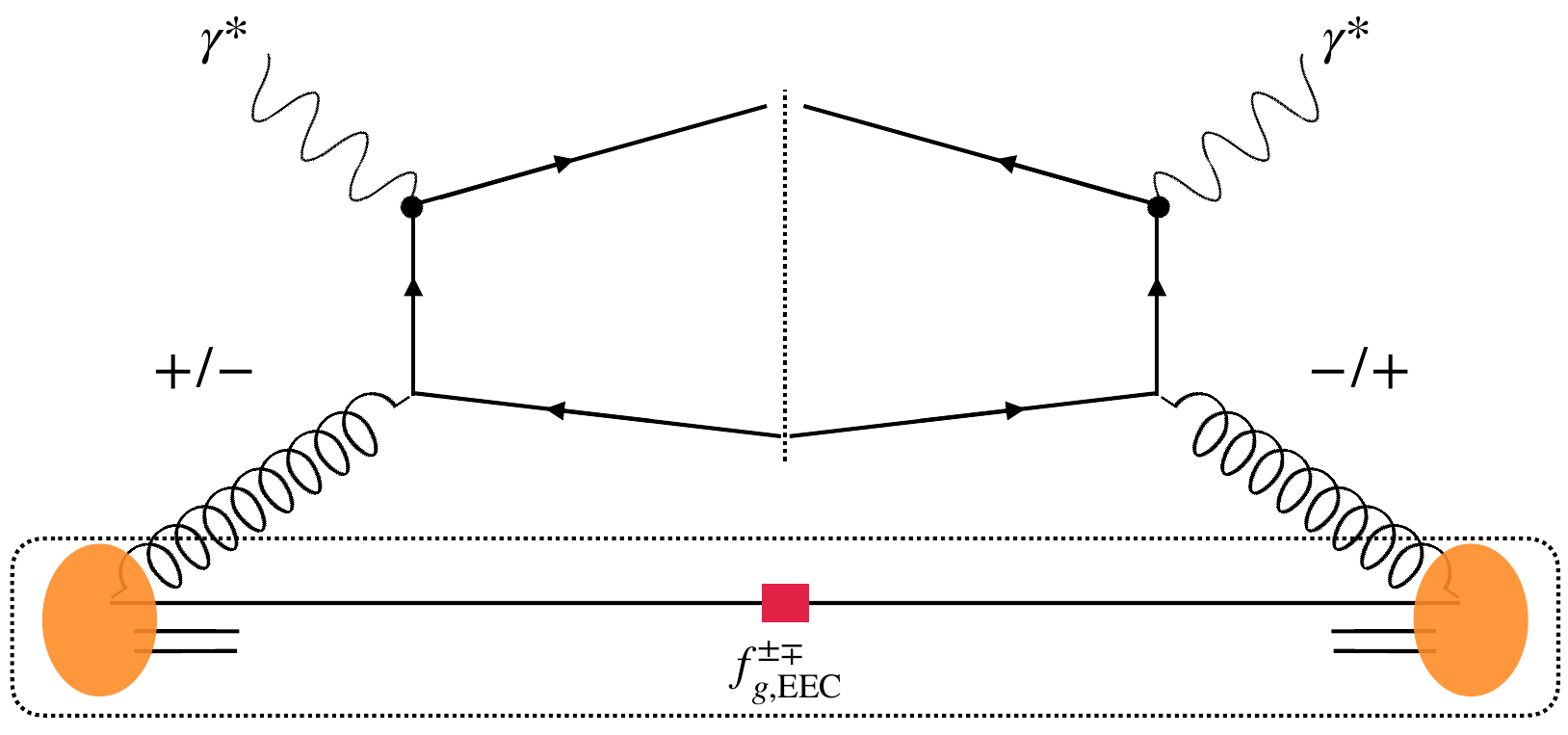} 
\caption{NEEC factorization theorem. Representative Feynman diagrams for ${\cal M}^\mu_g {\cal M}^{\nu\ast}_g$ at ${\cal O}(\alpha_s)$. }
  \label{fg:feyn}
 \end{center}
\end{figure}

The general form of the cross-section $\Sigma$ is given by 
\bea 
\Sigma 
= \frac{4\pi \alpha^2e_q^2}{Q^4} 
l_{\mu\nu} \Sigma^{\mu\nu} (x_B,\cos\theta_{a,b},\phi)  \,, 
\eea 
where $\alpha$ is the electrical fine structure constant and 
$l_{\mu\nu} =  g_{\mu\nu}(-2l\cdot l') + 4 l^\mu l^\nu$, where the Ward identity $q_\mu  \Sigma^{\mu\nu}  =0$ has been applied. Here 
$-2l\cdot l' = Q^2$ and 
$l^\mu = Q\frac{1+y}{y}(1,\frac{\sqrt{1+2y}}{1+y},0,\frac{y}{1+y})$. 
The $\Sigma^{\mu\nu}$ is the cross section for $\gamma^\ast P \to X$ with the energy correlators measured. When $\theta_a \ll \theta_b$, the  calculation of the $\Sigma^{\mu\nu}$ can be performed within the collinear NEEC factorization, closely follows~\cite{Cao:2023oef}, which gives
\bea\label{eq:sig-M}
\Sigma^{\mu\nu} &= &
\frac{y^2}{16\pi Q^2}  
 \int d\Phi_X
  {\cal M}^\mu_q {\cal M}^{\nu\dagger}_q f_{q,{\rm EEC}}(x,{\vec n}_a) \nn \\ 
&& \hspace{11.ex} + \,
 {\cal M}^\mu_{g,\alpha} 
 {\cal M}_{g,\beta}^{\nu \dagger} 
\, f^{\alpha\beta}_{g,{\rm EEC}}(x,{\vec n}_a)\,.
\eea 
The factorization theorem is illustrated in Fig.~\ref{fg:feyn}. 
 Here, $\Phi_X$ stands for the phase space of the final state partons, including the integration over the incoming parton momentum fraction $\int \frac{dx}{x}$,  with the energy $E_j/E_P$ weighting and the angle $\phi$ measurement in Eq.~(\ref{eq:e3c}) included.
 The $E_i/E_P$ and ${\vec n}_a$ measurements have been absorbed into the definition of the NEEC. ${\cal M}_i^\mu$ is the matrix element for the partonic $\gamma^\ast i \to jj+X$ production and can be calculated order by order in $\alpha_s$, whose leading-order (LO) contribution is illustrated in Fig.~\ref{fg:feyn}.  
The subscript $q$ ($g$) indicates the quark (gluon)-initiated partonic process. 
Here we have used the fact that in perturbative QCD (and also QED), the massless quark helicity is conserved,  therefore 
only the unpolarized quark NEEC $f_{q,{\rm EEC}}$ is involved.  

It can be shown that to all orders, Eq.~(\ref{eq:sig-M}) fulfills the general form such that, up to power corrections $\sim {\cal O}(\theta_a/\theta_b)$,
\bea\label{eq:sig-z} 
&& 
l_{\mu\nu}
\Sigma^{\mu\nu}
= \int \frac{dz}{z} \Bigg[
\sum_{i=q,g} \hat{H}_{i} (z,y,c_b) \frac{x_B}{z}f_{i,{\rm EEC}}\left(\frac{x_B}{z},\theta_a^2  \right) \nn \\ 
& + &
\frac{1}{2}\cos(2\phi)
\Delta\hat{H}_{g}(z,y,c_b ) \frac{x_B}{z} d_{g,{\rm EEC}} \left(\frac{x_B}{z},\theta_a^2 \right) \Bigg]\,,  
\eea 
 where we abbreviated $c_b = \cos\theta_b$ and suppressed the scale dependence. Here, $z \equiv \frac{x_B}{x}$ and the factor $\frac{x_B}{z}$ is originated from $E_j/E_P$. 
To understand how we get the form of Eq.~(\ref{eq:sig-z}), we first note that the parton momentum $p^\alpha = \frac{Q}{z}(1,0,0,1)$ that initiates the interaction is determined by $z$. Given that we are inclusive over the final state energy, then the partonic cross section $\hat{H}$ and $\Delta\hat{H}$ can only be functions of $z$, $y$, and $\vec{n}_b \approx \vec{d}$ in the small $\theta_a$ limit. The quark channel is un-polarized and hence $\phi$ independent. As for the gluon channel, given the tensor structure of $f_{g,{\rm EEC}}^{\alpha\beta}$ in Eq.~(\ref{eq:fgmunu}), any Lorentz structures of 
$\int d\Phi_X   l_{\mu\nu}  {\cal M}_{g,\alpha}^\mu 
  {\cal M}_{g,\beta}^{\nu,\dagger}$
constructed out of the longitudinal vectors vanishes when contracted with $f_{g,{\rm EEC}}^{\alpha\beta}$. Therefore, its non-vanishing contribution to $l_{\mu\nu}\Sigma^{\mu\nu}$ must require the all-order form  
\bea\label{eq:sig-M-trans} 
- g_T^{\alpha\beta} A(z,c_b)
+ \left(\frac{n_{b,T}^\alpha n_{b,T}^\beta }{n_{b,T}^2} - \frac{g_T^{\alpha\beta}}{2}\right) B(z,c_b)\,.
\eea 
 The form is determined by the reason that $\int d\Phi_X   l_{\mu\nu}  {\cal M}_{g,\alpha}^\mu 
  {\cal M}_{g,\beta}^{\nu,\dagger}$ is rotational covariant around the $z$-axis and can only be constructed out of $g_{\alpha\beta}$, $ p^\alpha$, $q^\alpha$ and $d^\alpha \approx n_b^\alpha$, while neither $p^\alpha$ nor $q^\alpha$ but only $n_b^\alpha$ acquires a transverse component. 
  Furthermore, the $\phi_b$ integration eliminates possible $\phi_b$ dependence within $A$ and $B$. 
  Now contracting Eq.~(\ref{eq:sig-M-trans}) with $f_{g,{\rm EEC}}^{\alpha\beta}$, we arrive at the final form in Eq.~(\ref{eq:sig-z}), where the unpolarized gluon contribution comes from the contraction of the $-g_{T}^{\alpha\beta}$ structures, and the $\cos(2\phi)$ from the $B$ term. Here, we have applied $\phi = \phi_d - \phi_a \approx \phi_b - \phi_a$ in the $\theta_a\to 0$ limit.


We conclude from Eq.~(\ref{eq:sig-z}) that the NEEC-based measurement provides a unique chance to probe the linearly polarized gluons since
\begin{itemize}[leftmargin=*]
\item Eq.~(\ref{eq:sig-z}), ~(\ref{eq:fgmunu}) and~(\ref{eq:sig-M-trans}) hold to all-orders with all radiation effects such as parton shower being taken into account. 
Therefore, unlike the TMDs, to all orders, the NEEC probe involves only one azimuthal structure $\cos(2\phi)$, due to the absence of soft radiations and hence no cross-talk between ${\cal M}_\mu{\cal M}^\dagger_\nu$ and $f^{\mu\nu}_{\rm EEC}$. Each of the ${\cal M}_\mu{\cal M}^\dagger_\nu$
and $f^{\mu\nu}_{\rm EEC}$ can only depend on one of the azimuthal angles ($\phi_a$ or $\phi_d \approx \phi_b$). Therefore, $\phi$ enters only through the tensor structure in Eq.~(\ref{eq:fgmunu}) and Eq.~(\ref{eq:sig-M-trans}), which uniquely determines the $\cos(2\phi)$ asymmetry. 

In contrast, the TMD soft radiation with momentum $k$ could simultaneously connect all directions, for instance, both the proton $P$ and the leading jet $P_J$ in the dijet process, and thus depends on all azimuthal angles. On that account, additional azimuthal dependence due to the eikonal factor $  \frac{1}{k\cdot P k\cdot P_J} \propto 
\frac{1}{\dots + \cos(\phi)} \to \sum_n c_n \cos(n\phi)$~\cite{Hatta:2020bgy,Hatta:2021jcd,Caucal:2023fsf}, contaminates the naive $\cos(2\phi)$ expectation. 

In this sense, the NEEC is a cleaner probe of the rotating gluons inside the nucleon target.

\item 
Furthermore, the NEEC factorization in Eq.~(\ref{eq:sig-z}) suffers no Sudakov suppression~\cite{Liu:2022wop,Liu:2023aqb,Cao:2023oef} in the non-perturbative signal region when $\theta_a Q\sim {\cal O}(\Lambda_{\rm QCD})$, which is quite different from the TMD case. On the contrary, the region is enhanced by the DGLAP evolution of the NEEC~\cite{Cao:2023oef}.


\item The energy flow measurement $E(\vec{n}_b)$ in the central region can be replaced by a jet constructed using a standard jet algorithm. The precision of this measurement can be improved by using the tracking information~\cite{Jaarsma:2022kdd,Li:2021zcf,Jaarsma:2023ell,Lee:2023xzv}. To further enhance the sensitivity to the gluon $f_{\rm EEC}$, we can tag the heavy quark species (charm/$b$-energy flow) for $E({\vec n}_b)$ measurement~\cite{Craft:2022kdo,Andres:2023ymw}.


\end{itemize}


\textbf{\textit{Numerics.}}
Now we present numerical studies. Our main objective of this study is to examine the all-order $\cos(2\phi)$ structure in Eq.~(\ref{eq:sig-z}), through a perturbative calculation at higher orders. 
We use the \texttt{nlojet++}~\cite{Nagy:2001xb} to generate tri-jet production in DIS at NLO (${\cal O}(\alpha_s^2+\alpha_s^3)$, up to four jets). Since the exact NEEC is not known, we model it by restricting $0.005<\theta_a < 0.02$. 
In this calculation, the strong coupling constant $\alpha_s$ is evaluated at $Q^2$ and $\alpha = 1/128.0$. 
Through our numerical study, we aim to provide a non-trivial test of the all-order factorization structure derived in Eq.~(\ref{eq:sig-z}), offering an initial insight into what could be expected from the measurement.

 \begin{figure}[htbp]
  \begin{center}
   \includegraphics[scale=0.42]{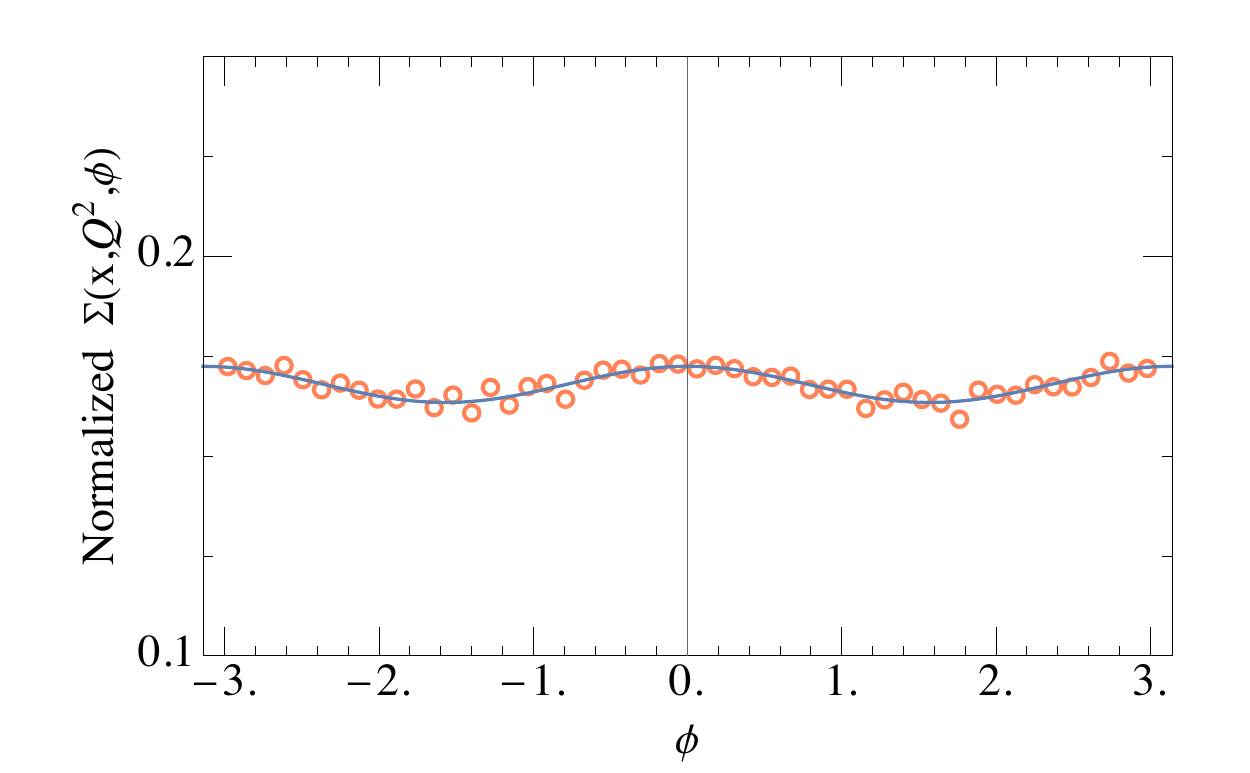} 
\caption{$\cos(2\phi)$ asymmetry at ${\cal O}(\alpha_s^2+\alpha_s^3)$ in the Breit frame.}
  \label{fg:breit}
 \end{center}
\end{figure}

In Fig.~\ref{fg:breit}, we show the 
 normalized \texttt{nlojet++} $\phi$ distribution result (in circle dots) at NLO ${\cal O}(\alpha_s^2+\alpha_s^3)$ in the Breit frame for 
$E_l = 18\, {\rm GeV}$,
$E_P = 275\, {\rm GeV}$, $Q^2 = 100\,{\rm GeV}^2$. We choose $x_B = 0.01$. We set $1.0<\theta_b < 1.5$.  We fit the un-normalized distribution with $a+b\cos(2\phi)$ (solid curve) to observe an excellent $\cos(2\phi)$ asymmetry to agree with our expectation from Eq.~(\ref{eq:sig-z}). 
Since both loop corrections and real emission up to 4 jet contributions are involved at this order, Fig.~\ref{fg:breit} acts as a highly non-trivial test of the all-order formalism we derived in Eq.~(\ref{eq:sig-z}). 

  \begin{figure}[htbp]
  \begin{center}
      \includegraphics[scale=0.42]{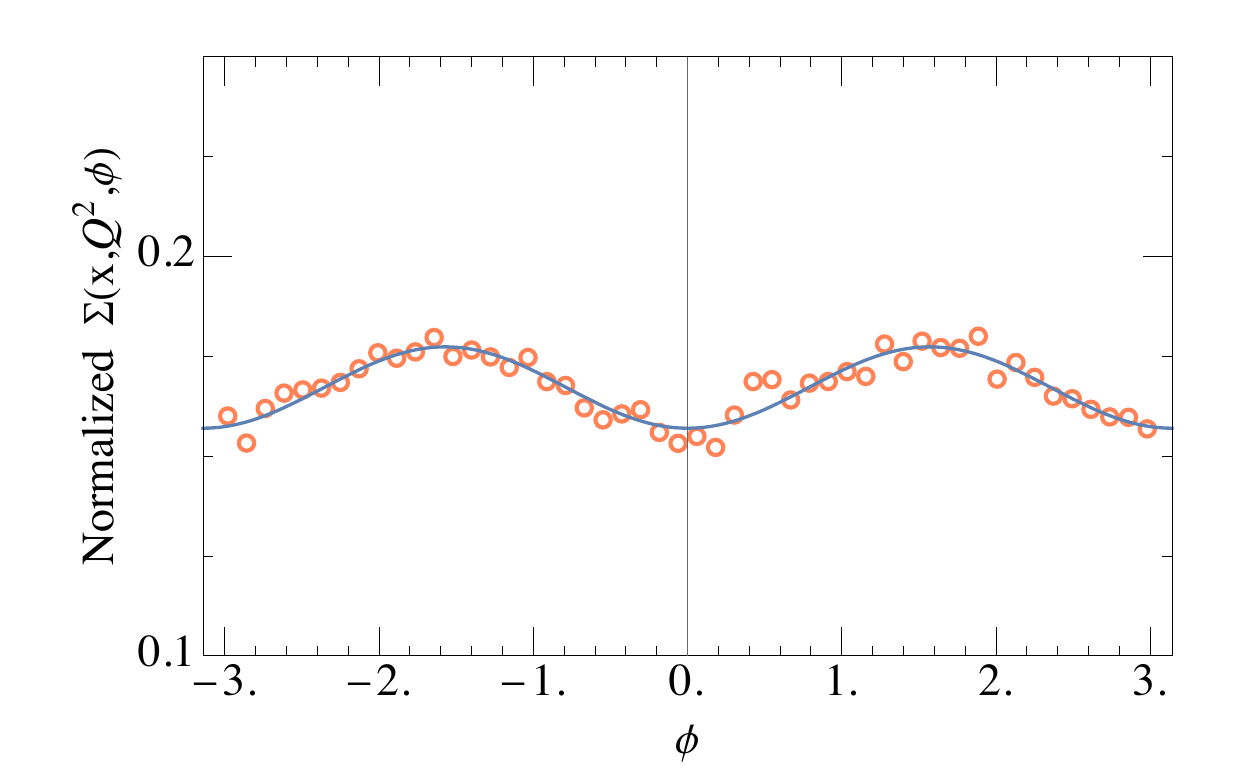} 
\caption{$\cos(2\phi)$ asymmetry in the squeezed limit, at  ${\cal O}(\alpha_s^2+\alpha_s^3)$ in the Breit frame}
  \label{fg:breit-squeeze}
 \end{center}
\end{figure}
In this calculation, we find the asymmetry induced by the linearly polarized gluon is $\Delta \hat{H}_g d_{g,\rm EEC}/\hat{H}_if_{i,{\rm EEC}} \sim b/a \approx~2.69\>\%$, as can be seen from Eq.~(\ref{eq:sig-z}). 
Here, $\Delta \hat{H}_g$ and $\hat{H}$ can be calculated perturbatively.  
Therefore, in reality, measuring the azimuthal asymmetry could tell directly the ``amount" of the linearly polarized gluons within the nucleon, once the unpolarized $f_{\rm EEC}$ is measured. 
We note that there could be logarithmic $\ln\theta_a$ correction to the normalized distribution showed in Fig.~\ref{fg:breit} which can be resummed through the evolution of $f_{i,{\rm EEC}}$ and $d_{g,{\rm EEC}}$ following the strategy in~\cite{Cao:2023oef}. However, resummation does not change the tensor structure in Eq.~(\ref{eq:fgmunu}) and thus leaves the $\cos(2\phi)$ pattern unaffected. Meanwhile, in this fixed order simulation, the strong coupling constant $\alpha_s$ is evaluated at $Q$ which will underestimate the size of the non-perturbative contribution. 
Nevertheless, since the primary objective of this study is to examine
the $\cos(2\phi)$ structure rather than determine the exact size of $d_{g,{\rm EEC}}$ (which in any case should be determined by future experiments), we have chosen to leave the $\ln\theta_a$ resummation for a subsequent study. We emphasize that Fig.~\ref{fg:breit} encompasses all channels, while, in practical measurements, the heavy flavor tagging for the energy deposition in the central region~\cite{Craft:2022kdo} could enhance the sensitivity to the gluon channel and improve the significance of the observed azimuthal asymmetry. 

Finally, we delve into the squeezed limit, defined as $\theta_a \ll \theta_b \ll 1$. We require $0.1 <\theta_b < 0.3$, $x_B=0.03$, and the results for the NLO normalized cases are shown in Fig.~\ref{fg:breit-squeeze}. Once again, we observe the presence of azimuthal asymmetry of size $b/a \approx -6.09\%$ 
indicating the persistence of the $\cos(2\phi)$ structure. We note that the sign of the asymmetry flips in Fig.~\ref{fg:breit} and Fig.~\ref{fg:breit-squeeze}, due to different kinematics regimes and the significant modification of large logarithms $\ln\theta_b$. We will carry out a detailed study in a future publication.

\textbf{\textit{Conclusion.}} Our findings indicate that it is possible to directly investigate the linearly polarized gluons through the observation of helicity-dependent NEEC in the DIS process. This method measures the energy deposition asymmetry in the calorimeter, which arises from the spinning gluon confined inside the nucleon and manifests as a $\cos(2\phi)$ correlation. Importantly, the $\cos(2\phi)$ signature is preserved by rotational symmetry and holds at all orders. Consequently, the shape of the asymmetry remains robust against radiation/parton shower contamination and free of Sudakov suppression.  The size of the asymmetry is to be determined by future experimental analysis and will provide us with a unique opportunity to determine the significance of gluon polarization at the current and future electron-ion facilities. Moreover, the absence of a polarized beam requirement suggests that it may be feasible to experimentally verify the factorization formalism using the available HERA data. Looking ahead, we plan to present the evolution of the helicity-dependent NEEC, as outlined in~\cite{Cao:2023oef}, to make all-order predictions for the azimuthal distribution. However, we do not anticipate any qualitative modifications to the results presented in this study.  


\begin{acknowledgments}


\textbf{\textit{Acknowledgement.}} We are grateful to Dingyu Shao and Jian Zhou for their useful discussions.
This work is supported by the Natural Science Foundation of China under contract No.~12175016 (X.~L.~L. and X.~L.), No.~11975200 (H.~X.~Z.) and the Office of Science of the U.S. Department of Energy under Contract No. DE-AC02-05CH11231 (F.Y.). 

 \end{acknowledgments}

\bibliographystyle{h-physrev}   
\bibliography{refs}


\end{document}